\documentclass[12pt]{article}
\usepackage{amsfonts,amsmath,amssymb,epsf}
\usepackage{graphicx,color}
\newcommand{\be}{\begin{equation}}
\newcommand{\ba}{\begin{eqnarray}}
\newcommand{\ea}{\end{eqnarray}}
\newcommand{\ee}{\end{equation}}

\begin{document}

\begin{flushright} 
June 2008  \\
KUNS-2146\\
WIS/13/08-Jun-DPP
\end{flushright} 

\vspace{0.1cm}

\begin{center}
  {\LARGE
   On Matrix Model Formulations\\
 of Noncommutative Yang-Mills Theories
  }
\end{center}
\vspace{0.1cm}
\begin{center}

         Tatsuo A{\sc zeyanagi}$^{a}$\footnote
           {
E-mail address : aze@gauge.scphys.kyoto-u.ac.jp},  
          Masanori H{\sc anada}$^{b}$\footnote
           {
E-mail address : masanori.hanada@weizmann.ac.il} and   
          Tomoyoshi H{\sc irata}$^{a}$\footnote
           {
E-mail address : hirata@gauge.scphys.kyoto-u.ac.jp}

\vspace{0.3cm}

$^a$           
{\it Department of Physics, Kyoto University,\\
Kyoto 606-8502, Japan}\\

$^b$           
{\it Department of Particle Physics, Weizmann Institute of Science\\
     Rehovot 76100, Israel }\\

\end{center}

\vspace{1.5cm}

\begin{center}
  {\bf abstract}
\end{center}

We study stability of noncommutative spaces 
in matrix models and discuss the continuum limit 
which leads to noncommutative Yang-Mills theories (NCYM). 
It turns out that most of noncommutative spaces in bosonic models are 
unstable.
This indicates perturbative 
instability of fuzzy ${\mathbb R}^D$ pointed out by Van Raamsdonk and Armoni et al.
persists to nonperturbative level in these cases. 
In this sense, these bosonic NCYM are not well-defined, or at least 
their matrix model formulations studied in this paper do not work. 
We also show that noncommutative backgrounds are stable in 
a supersymmetric matrix model deformed by a cubic Myers term, 
though the deformation itself breaks supersymmetry.

\newpage
\newpage
\section{Introduction}
\hspace{0.51cm}
Yang-Mills theory on a noncommutative space 
(noncommutative Yang-Mills theory, or simply NCYM) has attracted much interest in theoretical physics. 
It appears as an effective theory of string theory or 
its matrix models around certain flux backgrounds \cite{SW99}\cite{BFSS96,IKKT96}
\cite{CDS97,AIIKKT99,Li96}. 
NCYM contains some interesting physical properties like spacetime 
uncertainty and peculiar solitonic solutions \cite{GMS00}.
We also notice that it naturally contains gravity (for recent progress, 
see e.g. \cite{GravityInNCYM,SteinackerMatter}). 
To understand the nonperturbative aspects of NCYM better, 
we need the nonperturbative formulation of it. 
Matrix models are expected to be the most promising approach. 
Using a matrix model, NCYM is realized as an effective theory of a matrix model around a certain background. 
However, such backgrounds are unstable for some cases and whether the theories are well defined or not is a nontrivial question. 
It is well defined only when the backgrounds are stable. 
In this note, we will discuss stability of noncommutative 
spaces and argue what kinds of NCYM can be realized using matrix models. 

Realization of NCYM in matrix models is of interest also from 
{\it emergent geometry} point of view.  
The origin of this concept goes back to early 1980s.
The first example, as far as we know, is large-$N$ reduction 
\cite{EK82,QEK,GAO82} which claims that large-$N$ gauge theories are 
equivalent to their one point reduced models.
In these models, spacetime is embedded in gauge fields \cite{QEK,GAO82,EN82}. 
We can also find it in the context of quantum theory of gravity.
From this point of view, spacetime should emerge as a result of some dynamical mechanism. As nonperturbative formulations of string theory, 
various matrix models are proposed \cite{BFSS96,IKKT96} and, especially in 
IKKT matrix model \cite{IKKT96}, various interpretations are given to realize 
emergent geometry \cite{EmergentGeometryInIKKT,DifferentialOperator,SteinackerMatter}. 
This concept is also discussed in the context of AdS/CFT \cite{Maldacena97}\cite{Berenstein05}. 
   
NCYM is another example of emergent geometry. Let us briefly explain how it shows up and what kind of double scaling limit is necessary. 
We only consider NCYM on a flat noncommutative space and 
mainly take the continuum limit in which 
the noncommutativity parameter $\theta$ is fixed.  
We set the gauge group to be $U(1)$ unless otherwise mentioned but 
generalization to $U(n)$ is straightforward. 

For concreteness, let us consider zero-dimensional $SU(N)$ matrix models  
with a twisted boundary condition \cite{GAO82} or a Myers term added \cite{Myers99}.
For these models, it is known that compact noncommutative spaces like 
fuzzy spheres are classical solutions.
Once we fix $\theta$, volume of the space and the UV cutoff are related to the matrix size $N$. Therefore, the gauge coupling $g_{NC}$ runs with $N$. 
Strictly speaking, renormalizability of NCYM is a subtle problem.  
In principle, using numerical simulations, 
the scaling is determined nonperturbatively so that some renormalization condition is satisfied.
For example, in \cite{BHN02}, $D=2$ case is discussed and 
renormalization is performed so that 
the expectation value of the Wilson loop with the same area in physical unit is kept fixed. This result is equivalent to the one for the one-loop calculation.
In principle, we can similarly perform renormalization for the case of $D=4$, however, it is hard with current numerical resources. Therefore, we rely on the one-loop 
calculation for this case \cite{MRS99}. 
It is known that for Non-Abelian gauge theory, 
the scaling of the gauge coupling turns out to be the same as that of the commutative case. 
On the other hand, the case of Abelian gauge theory is 
extremely different and it is known that the beta function is the same as that of Non-Abelian gauge theory on commutative space.
That is, for NCYM, Abelian gauge theory is also 
asymptotically free as a result of the existence of non-planar diagrams.  
 
In order for NCYM to be well-defined, noncommutative spaces must be stable in this double scaling limit.     
However, in some cases $g_{NC}$ runs into a region where the space is not stable any more. 
We show that this is the case for most of bosonic models.
Therefore, as suspected for a long time \cite{VanRaamsdonk01}, 
NCYM on fuzzy ${\mathbb R}^D$ is not well-defined nonperturbatively  
(At least matrix model formulations discussed in this paper do not work).
Here we also notice that $D=2$ pure NCYM is only one exception that we have found in this paper.  
In other words, NCYM describes a wrong vacuum and hence noncommutative 
spacetime is not an emergent background in this case. 
This is not necessarily a negative conclusion - we can say that 
NCYM correctly describes gravitational instability.   

On the other hand, once supersymmetry is introduced we can expect that 
fuzzy spaces are stabilized because of the BPS nature and 
noncommutative super Yang-Mills theory (NCSYM) on fuzzy $\mathbb{R}^4$ 
is realized.
In order to formulate NCSYM on fuzzy $\mathbb{R}^4$, 
we add a cubic Myers term to the usual IKKT-like matrix models. 
One thing we notice here is the fact that this models themselves do not have
supersymmetry but it recovers in the double scaling limit.
 
Organization of this paper is as follows. 
In \S\ref{sec:bosonic} we study bosonic matrix models to formulate 
bosonic NCYM. 
We firstly discuss the twisted Eguchi-Kawai model \cite{GAO82}  
and explain that we cannot formulate $D=4$ pure NCYM \cite{AHHI07} 
while we can formulate  
$D=2$ pure NCYM. 
Next, we discuss bosonic analogues of IKKT matrix models 
with a cubic Myers term and analyze the stability of solutions like fuzzy spheres.
We show that we cannot formulate $D=4$ 
and $D=2$ NCYM with adjoint scalars  
using this formulation.
We also demonstrate that pure $D=2$ NCYM can be realized by adding a potential 
term to an adjoint scalar. 
In the end of this section,
we comments on other scaling limits like {\it commutative} limit. 
In \S\ref{sec:NCSYM} we study approximately  
supersymmetric matrix models with a cubic Myers term 
to formulate NCSYM and show that the approximate supersymmetry stabilizes 
fuzzy spaces.

\section{Bosonic matrix models and bosonic NCYM \\on fuzzy ${\mathbb R}^D$}
\label{sec:bosonic}
\hspace{0.51cm}
In this section, we study bosonic matrix models and their 
double scaling limit which leads to bosonic NCYM on fuzzy ${\mathbb R}^D$. 
In \S\ref{sec:TEK} we briefly review the twisted Eguchi-Kawai model 
(TEK) \cite{GAO82} and 
discuss the stability of the ground state \cite{AHHI07,TV06}.
In \S\ref{sec:TEKNCYM} we explain the formulation 
of NCYM using TEK \cite{AIIKKT99,AMNS99} and explain the double scaling limit. 
It turns out that NCYM on fuzzy ${\mathbb R}^4$ cannot be 
realized using it \cite{AHHI07}.   
In \S\ref{sec:cubic term} we introduce 
bosonic analogue to IKKT matrix model with a cubic Myers term, 
which has fuzzy $S^2\times S^2$ as a classical solution. 
We show that this background is unstable in the double scaling limit. 
Discussion in this subsection applies also to other deformations with a cubic Myers term. 
In \S\ref{otherlimit} we study other possible limits including commutative limit. 
\subsection{Twisted Eguchi-Kawai model}\label{sec:TEK}
\hspace{0.51cm}
Twisted Eguchi-Kawai model (TEK) \cite{GAO82} 
is a unitary matrix model
defined by the action
\begin{eqnarray}
S_{TEK}=-\beta N\sum_{\mu\neq\nu}
        Z_{\mu\nu}Tr\left(U_\mu U_\nu U_\mu^\dagger U_\nu^\dagger\right),
\label{action:TEK}
\end{eqnarray}
where $U_{\mu}$ are $N\times N$ unitary matrices
with the Greek indices run from $1$ to $D$ 
and $\beta$ is the inverse of the 't Hooft coupling.
We mainly concentrate on the case of $D=4$.
We comment on the case of $D=2$ in the end of the next subsection where 
we discuss matrix formulation of NCYM on fuzzy ${\mathbb R}^2$.

The phase factors $Z_{\mu\nu}$ are defined by
\begin{eqnarray}
Z_{\mu\nu}=\exp\left(2\pi i n_{\mu\nu}/N\right),
\qquad
n_{\mu\nu}=-n_{\nu\mu}\in \mathbb{Z}_N.\nonumber
\end{eqnarray}
In this paper, we use the skew diagonal twist which is written 
as
\begin{eqnarray}
(n_{\mu\nu})
=
\left(\begin{array}{cc|cc}
 0 &  L &  0 & 0\\
-L &  0 &  0 & 0\\
\hline
 0 &  0 &  0 & L\\
 0 &  0 & -L & 0
\end{array}\right),
\label{skew-diagonal form 4d}
\end{eqnarray}
where $L=\sqrt{N}$ corresponds to the lattice size
\cite{GAO82}. 
There are other ways of twisting, but discussion is completely parallel and conclusion is the same as far as the double scaling limit which leads to NCYM is concerned.
  
In the weak coupling limit ($\beta\to\infty$), the path-integral is dominated by configurations with the minimum value of the action.
This configuration $U^{(0)}_\mu=\Gamma_{\mu}$ is called  
``twist eater'' and satisfies the 't Hooft algebra 
\begin{eqnarray}
\Gamma_{\mu}\Gamma_{\nu}=Z_{\nu\mu}\Gamma_{\nu}\Gamma_{\mu}. 
\label{'t Hooft algebra}
\end{eqnarray}
For the skew-diagonal twist, we can easily construct a 
twist eater configuration by introducing $L \times L$ 
``shift'' matrix $\hat{S}_L$ and ``clock''
matrix $\hat{C}_L$    
\begin{eqnarray}
\hat{S}_L=\left(
   \begin{array}{ccccc}
     0 & 1 & 0 & \cdots & 0 \\
     0 & 0 & 1 & \cdots & 0 \\
     \vdots & \vdots & \vdots & \ddots & \vdots \\
     0 & 0 & 0 & \cdots & 1 \\
     1 & 0 & 0 & \cdots & 0
   \end{array}
\right), \qquad
\hat{C}_L=\left(
   \begin{array}{ccccc}
     1 &   &   &   &   \\
     & e^{2\pi i/L} & & & \\
     & &   e^{2\pi i\cdot 2/L} & & \\
     & & & \ddots & \\
     & & & & e^{2\pi i(L-1)/L}
   \end{array}
\right).
\label{shift and clock}
\end{eqnarray}
These matrices satisfy 
\begin{eqnarray}
\hat{C}_L\hat{S}_L =  e^{-2\pi i/L}\hat{S}_L\hat{C}_L,
\label{little algebra}
\end{eqnarray}
and then we can construct a twist eater configuration for the 
above skew diagonal twist as 
\begin{eqnarray}
\Gamma_1=\hat{C}_L     \otimes \textbf{1}_L, \quad
\Gamma_2=\hat{S}_L     \otimes \textbf{1}_L, \nonumber\\
\Gamma_3=\textbf{1}_L \otimes \hat{C}_L,     \quad
\Gamma_4=\textbf{1}_L \otimes \hat{S}_L.
\label{vacuum for minimal skew diagonal twist 4d}
\end{eqnarray}
This twist eater configuration is nothing but fuzzy $T^4$ in the context of NCYM. 
We will explain the relation between fuzzy $T^4$ and fuzzy ${\mathbb R}^4$ 
when we use TEK as a potential nonperturbative formulation of NCYM on 
fuzzy ${\mathbb R}^4$ in the next subsection. 

In \cite{AHHI07}, it was shown, by Monte-Carlo study of TEK, 
that the configuration deviates from the $ \Gamma_\mu$ and the fuzzy torus collapses in a certain range of the 
inverse 't Hooft coupling $\beta$.
The upper boundary of this region scales as 
\begin{eqnarray}
\beta_c\simeq 0.0034N+0.25. 
\label{EQ:bc_L_minimal_sym}
\end{eqnarray}
We can estimate this behavior  easily and somehow roughly as follows. 
For simplicity, we assume that the fuzzy torus $U_{\mu}=\Gamma_{\mu}$ 
collapses to the identity configuration $U_{\mu}=\textbf{1}_N$.
The difference of energy between these configurations is  
\begin{equation}
\Delta S=S_{TEK}(U_\mu= \textbf{1}_N)-S_{TEK}(U_\mu= \Gamma_\mu)=
8\pi^2\beta N.
\label{EQ:energy difference generic}
\end{equation}
Far from the weak coupling limit, the system has quantum fluctuations.
Especially quantum fluctuations about twist-eater 
is known to be $O(N^2)$ \cite{HNT98}. 
Roughly expecting that the fuzzy torus collapses 
if the fluctuation around twist-eater configuration exceeds
the energy difference $\Delta S$,
we can estimate the critical point $\beta_c^L$ on which 
the torus begins to collapse as
\footnote{
In this paper, we often estimate the  
power of $N$ only and we use ``$A$ $\sim$ $B$'' (resp. ``$A$ $\lesssim$ $B$'') 
to represent that the order of $A$ is
equal to (resp. equal to or less than) that of $B$.}
\begin{equation}
\beta_c\sim N,
\end{equation}
which is consistent with the numerical results (\ref{EQ:bc_L_minimal_sym}). 
\subsection{TEK and NCYM on fuzzy ${\mathbb R}^D$}\label{sec:TEKNCYM}
\hspace{0.51cm}
TEK is a potential nonperturbative formulation of pure NCYM
on fuzzy ${\mathbb R}^D$.
\cite{AIIKKT99,AMNS99}.
In order to realize the formulation, we notice 
that fuzzy ${\mathbb R}^D$ is realized as a tangent space of fuzzy $T^D$. We can determine 
whether we can formulate the NCYM or not by analyzing the stability of the torus in the double scaling limit.   
Here we review the formulation of NCYM on 
${\mathbb R}^4$ using TEK and especially  
discuss the double scaling limit \cite{AHHI07} and 
the stability of the fuzzy $T^4$.
We also comment on the case of $D=2$.

By taking $U_\mu=e^{iaA_\mu}$, where $a$ corresponds to the lattice
spacing, and expanding the action of TEK (\ref{action:TEK}),
we have its continuum version as
\begin{eqnarray}
S_{TEK}=-\frac{1}{4g^2}\sum_{\mu\neq\nu}
         Tr\left([A_\mu,A_\nu]-i\theta_{\mu\nu}\right)^2,
\label{continuum TEK}
\end{eqnarray}
up to higher order terms in $a$, where 
\begin{eqnarray}
\theta_{\mu\nu}=\frac{2\pi n_{\mu\nu}}{Na^2},\quad
\frac{1}{4g^2}=a^4 \beta N.   
\end{eqnarray}
Then, by expanding the action around a classical solution 
(\ref{continuum TEK})
\begin{eqnarray}
A_\mu^{(0)}=\hat{p}_\mu, \qquad
[\hat{p}_\mu,\hat{p}_\nu]=i\theta_{\mu\nu},   
\end{eqnarray}
we obtain the $U(1)$ NCYM on fuzzy ${\mathbb R}^4$ as follows. 
Let us define the ``noncommutative coordinate''
$\hat{x}^\mu=\left(\theta^{-1}\right)^{\mu\nu}\hat{p}_\nu$.
Then we have 
\begin{eqnarray}
  [\hat{x}^\mu,\hat{x}^\nu]=-i(\theta^{-1})^{\mu\nu}. 
\end{eqnarray}
This commutation relation is the same as that of 
coordinates on fuzzy ${\mathbb R}^4$ with noncommutativity parameter
$\theta$, and hence functions of $\hat{x}$ can be mapped to 
functions on fuzzy ${\mathbb R}^4$. More precisely, we have the
following mapping rule:  
\begin{eqnarray}
  \begin{array}{ccc}
    f(\hat{x})=\sum_k\tilde{f}(k)e^{ik\hat{x}}
    &\leftrightarrow&
    f(x)=\sum_k\tilde{f}(k)e^{ikx}, 
    \\
    f(\hat{x})g(\hat{x})
    &\leftrightarrow&
    f(x)\star g(x), \\
    i[\hat{p}_\mu,\ \cdot\ ]
    &\leftrightarrow&
    \partial_\mu, \\
    Tr
    &\leftrightarrow&
    \frac{\sqrt{\det\theta}}{4\pi^2}\int d^4x,  
  \end{array}
\end{eqnarray}
where $\star$ represents the noncommutative product, 
\begin{eqnarray}
  f(x)\star g(x)
  =
  f(x)\exp\left(-\frac{i}{2}
    \overset{\leftarrow}{\partial}_\mu
    (\theta^{-1})^{\mu\nu}
    \overset{\rightarrow}{\partial}_\nu
  \right) g(x),
\end{eqnarray}
and we obtain $U(1)$ NCYM with coupling constant 
\begin{equation}
g_{NC}^2=4\pi^2g^2/\sqrt{\det\theta}.
\end{equation}
In order to keep the noncommutative scale $\theta$ finite, we should take the
double scaling limit with
\begin{eqnarray}
a^{-1}\sim \Lambda\sim N^{1/4}. 
\end{eqnarray}
 
As we have explained the identification to formulate pure NCYM 
using TEK, we next determine the double scaling limit explicitly 
and discuss the stability of the fuzzy $T^4$.
The one-loop beta function for $D=4$ $U(1)$ NCYM is given by \cite{MRS99}
\begin{eqnarray}
\beta_{1-loop}(g_{NC})=-\frac{1}{(4\pi)^2}\frac{11}{3}g_{NC}^3+O(g_{NC}^5). 
\end{eqnarray}
Therefore, the inverse 't Hooft coupling $\beta$ scales as
\begin{eqnarray}
\beta \sim \frac{1}{g_{NC}^2} \sim \log\Lambda\sim\log N. 
\label{scaling_TEK}
\end{eqnarray}
Since we know that the torus collapses below the critical point $\beta_c$ which scales as 
(\ref{EQ:bc_L_minimal_sym}), we can see that the 
fuzzy $T^4$ collapses in the double scaling limit. 
Therefore we finally see that we cannot formulate 
$D=4$ pure NCYM using TEK. 

Before closing this subsection we comments on the results for $D=2$.
In this case, (\ref{skew-diagonal form 4d}) and 
(\ref{vacuum for minimal skew diagonal twist 4d}) are replaced by 
\begin{eqnarray}
(n_{\mu\nu})
=
\left(\begin{array}{cc}
 0 &  1 \\
-1 &  0 \\
\end{array}\right).
\label{skew-diagonal form 2d}
\end{eqnarray}
and 
\begin{eqnarray}
\Gamma_1=\hat{C}_N , \quad
\Gamma_2=\hat{S}_N. 
\label{vacuum for minimal skew diagonal twist 2d}
\end{eqnarray}
This corresponds to a fuzzy $T^2$.
In order to take the double scaling limit 
with the noncommutative parameter $\theta\sim (a^2N)^{-1}$ fixed, 
we must scale the lattice spacing as $a\sim N^{-1/2}$. 
For $D=2$, the double scaling limit 
is determined by numerical simulations as \cite{BHN02}
\begin{eqnarray}
\beta\sim N. 
\end{eqnarray}
In this scaling, $g^2\sim g_{NC}^2$ does not run, which is consistent with 
one-loop beta function for $D=2$.  
Since we know that fuzzy $T^2$ does not collapse in $D=2$ TEK 
(This case is exceptional because there are no physical degrees of freedom.), 
we see that we can formulate $D=2$ pure NCYM on fuzzy ${\mathbb R}^2$. 
For detailed simulations and renormalizability see \cite{BHN02}. 
\subsection{Matrix model with a cubic Myers term}\label{sec:cubic term}
\hspace{0.51cm}
In this section, we use an bosonic analog to IKKT-type matrix model with a cubic Myers term.
More concretely, we consider a matrix model which 
has $S^2\times S^2$ as a classical solution by choosing the cubic term coupling appropriately.
Then we discuss stability of fuzzy $S^2\times S^2$.
Although we use the specific solution,  
this argument itself can be applied to the case of 
other compact noncommutative 
manifolds like fuzzy $S^2$ and fuzzy $\mathbb{C}P^2$.
\footnote{
The argument below can be parallelly applied to the case of
fuzzy $S^2$ and fuzzy $\mathbb{C}P^2$ because  
they are 
classical solutions of (\ref{matrix model with cubic term}) 
with  $f^{\mu\nu\rho}$ appropriately chosen.  
Fuzzy $S^4$, however, is a classical solution to a bosonic matrix model 
with a quintic Myers term. In this case, perturbative calculation is 
 not valid. 
In \cite{ABNN04S4}, it is numerically 
shown that $S^4$ is unstable in a bosonic matrix model.}  

Let us start with the $d=6$ bosonic analog to IKKT model with a cubic 
Myers term. The
action is written as 
\begin{eqnarray}
S=\frac{1}{g^2}Tr\left(
-\frac{1}{4}[A_\mu,A_\nu]^2
+
\frac{2i}{3}\alpha f^{\mu\nu\rho}A_\mu A_\nu A_\rho
\right), 
\label{matrix model with cubic term}
\end{eqnarray}
where $A_{\mu}$ is $N\times N$ hermitian matrix and the Greek 
indices run form $1$ to $6$. 
In the cubic term,
$f^{\mu\nu\rho}$ is the structure constant of $SU(2)\times SU(2)$ 
and $\alpha$ is a constant which characterizes the radii of fuzzy spheres. 
We choose the totally anti-symmetric tensor $f^{\mu\nu\rho}$ such that the only nonzero components are 
$f^{123}=f^{456}=1$ and their permutations. 

The equation of motion for this model is 
\begin{eqnarray}
[[A_\mu,A_\nu],A_\nu]+2i\alpha f^{\mu\nu\rho}A_\nu A_\rho=0. 
\end{eqnarray}
A classical solution called fuzzy $S^2\times S^2$ is given by
\begin{eqnarray}
A^{(0)}_\mu=\alpha J_\mu, 
\end{eqnarray}
where $J_\mu$ is a generator of $SU(2)\times SU(2)$ which satisfies 
\begin{eqnarray}
[J_\mu,J_\nu]=if^{\mu\nu\rho}J_\rho. 
\end{eqnarray}
$J_\mu$ can be expressed as\footnote{
We can also combine generators with different spins, 
but the argument does not change qualitatively. 
}
\begin{eqnarray}
J_{1,2,3}=J^{(s)}_{1,2,3}\otimes\textbf{1}_{2s+1}, 
\qquad
J_{4,5,6}=\textbf{1}_{2s+1}\otimes J^{(s)}_{1,2,3}
\label{classical solution_S2*S2}
\end{eqnarray}
where $J^{(s)}$ is the spin-$s$ generator and $N=(2s+1)^2$.  
In this case two fuzzy spheres have the same radius and 
square of the radius $R$ of the fuzzy sphere is given by
\begin{eqnarray}
 R^2=\displaystyle{\sum_{i=1}^{3}}\left(A_{i}^{(0)}\right)^2=\alpha^2s(s+1).
\end{eqnarray}

 Expanding the matrix model (\ref{matrix model with cubic term}) about 
(\ref{classical solution_S2*S2}), we obtain NCYM on fuzzy $S^2\times S^2$ coupled to two adjoint scalars.  
By zooming-up the north pole, i.e. considering only states with $J_3\sim J_6\sim s$, we formally obtain 
NCYM on fuzzy $\mathbb{R}^4$ with two adjoint scalars, which originate from transverse directions of the fuzzy $S^2 \times S^2$. Because 
\begin{eqnarray}
[A^{(0)}_1, A^{(0)}_2]=[A^{(0)}_4, A^{(0)}_5]=i\alpha^2 J_3\sim i\alpha^2 s, 
\end{eqnarray}
the noncommutativity parameter $\theta$ is 
\begin{eqnarray}
\theta\sim \alpha^2 s\sim \alpha^2\sqrt{N}. 
\end{eqnarray}
In order to keep $\theta$ fixed, we must scale $\alpha$ as
\begin{eqnarray}
\alpha\sim N^{-1/4},
\label{eq:scaling_of_alpha}
\end{eqnarray} 
and therefore the momentum cutoff scales as  
\begin{eqnarray}
\Lambda \sim \alpha s \sim N^{1/4}. 
\end{eqnarray}
As a result, in order to take the continuum limit with $\theta$ fixed, 
we have to scale $g^2$ as \cite{MRS99}
\begin{eqnarray}
g^{-2}=\frac{4\pi^2}{\theta^2} g_{NC}^{-2}\sim \log\Lambda\sim\log N. 
\label{eq:scaling_of_g}
\end{eqnarray}
 
We can easily see that fuzzy $S^2\times S^2$ collapses when
$\frac{1}{g^2}\lesssim N$, 
because energy difference between fuzzy $S^2\times S^2$ and $A_\mu=0$ is of order 
$\frac{\alpha^4 N^2}{g^2}\sim\frac{N}{g^2}$, while quantum fluctuations are of order 
$N^2$.  
Therefore, fuzzy $S^2\times S^2$ collapses when we take the double scaling limit
(\ref{eq:scaling_of_alpha}), (\ref{eq:scaling_of_g}) and then
we cannot take the continuum limit. 

This bound was derived more rigorously using Monte-Carlo simulation.  
Interestingly, this bound can also be derived 
through perturbative calculations of the matrix model \cite{InstabilityOfFuzzySphere}. 
Firstly, notice that eigenvalues are concentrated around the origin after the collapse of fuzzy sphere.
This can be confirmed by numerical simulations.
Then it is reasonable to assume that, in the perturbative analysis, 
such instability can be detected by 
considering only ``rescaled fuzzy sphere" 
$A_\mu^{{\rm rescaled}}=A_\mu^{(0)}\times const$ and by calculating its free energy as a function of the radius. 
At large enough coupling, there is a minimum to the free energy, which indicates that the background is stable. However, below some critical 
point, the minimum disappears and we can expect that the background is 
not stable any more. This is actually the case and the critical value 
obtained in this way agrees with numerical result very accurately.  
For details, see \cite{InstabilityOfFuzzySphere}.
In the next section, 
We assume the validity of the perturbative calculation and use it to justify the matrix formulation of supersymmetric noncommutative Yang-Mills theory.  

Here we comment on results for the formulation of 
$D=2$ NCYM with an adjoint scalar.  
Let us take $f^{\mu\nu\rho}$ to be the structure constant $\epsilon_{\mu\nu\rho}$
of $SU(2)$ where the Greek indices run form $1$ to $3$.  
As a classical solution of this matrix model
we can obtain fuzzy $S^2$. 
By zooming up the north pole as we did above, we obtain 
 the NCYM on fuzzy ${\mathbb R}^2$. 
In order for the noncommutativity parameter to be fixed, 
we have to scale the coupling constant for the cubic Myers term as
\begin{eqnarray}  
\alpha\sim N^{-1/2}.
\end{eqnarray} 
Because 
potential difference between the fuzzy $S^2$ and $A_{\mu}=0$
is of order $\frac{\alpha^4N^3}{g^2}\sim\frac{N}{g^2}$ 
while one-loop fluctuation is of order $N^2$, fuzzy $S^2$ collapses when 
$\frac{1}{g^2}\lesssim N$.
On the other hand, we can see the gauge coupling constant 
$g^2$ does not run similarly to the case of \S\ref{sec:TEKNCYM}. 
Therefore, we cannot take the continuum limit as 
$D=2$ NCYM with an adjoint scalar.
 
\subsubsection{Adding potential terms for adjoint scalars}\label{sec:SteinackerModel}
\hspace{0.51cm}
In \cite{Steinacker03}, another 
matrix model formulation of NCYM is introduced.
This matrix model has fuzzy $S^2$ as a classical solution. In the original paper above,    
the commutative limit $\theta\to\infty$ was studied. 
In this section, we rather discuss the double scaling limit 
with $\theta$ fixed and see whether we can use this matrix model to formulate NCYM.

For this model the action is given by  
\begin{eqnarray}
S=\frac{1}{4g^2}Tr\left\{
\left(
\alpha A_i+i\epsilon_{ijk}A_jA_k
\right)^2
+
\left(
A_i^2-\frac{\alpha^2}{4}(N^2-1)
\right)^2
\right\}. 
\label{steinacker's action}
\end{eqnarray}
By expanding the action about a classical solution 
\begin{eqnarray}
A_i=\alpha J_i,  
\label{steinacker_fuzzyS2}
\end{eqnarray}
where $J_i$ are $SU(2)$ generators with spin $s=\frac{N-1}{2}$, 
NCYM on fuzzy $S^2$ is realized. 
(The second term in (\ref{steinacker's action}) gives potential for adjoint scalar.)  
To take a continuum limit with fixed noncommutativity parameter, 
we should take large-$N$ limit with $g^2$ fixed and $\alpha\sim\frac{1}{\sqrt{N}}$.    

However, we can easily see that this background can collapse to a point e.g. 
\begin{eqnarray}
A_1=A_2=0, 
\qquad
A_3=\frac{\alpha \sqrt{N^2-1}}{2}. 
\label{steinacker_collapse}
\end{eqnarray}
We can easily see the difference of tree-level potential at 
(\ref{steinacker_fuzzyS2}) and a (\ref{steinacker_collapse}) is 
of order $\frac{\alpha^4 N^3}{g^2}$, while 
quantum fluctuations are of order $N^2$ in the double scaling limit. 
Therefore we can see that the critical coupling is $\frac{1}{g_c^2}\sim N$ 
and fuzzy $S^2$ collapses in the limit with $\frac{1}{g^2}\lesssim \frac{1}{g_c^2}$.   

In \cite{OY06} slightly generalized version of (\ref{steinacker's action}), 
\begin{eqnarray}
S=
N\ Tr\left\{
-\frac{1}{4}[X_i,X_j]^2
+
\frac{2i\rho}{3}\epsilon^{ijk}X_iX_jX_k
-
m^2\rho^2 X_i^2
+
\frac{2m^2}{N^2-1}(X_i^2)^2
\right\} ,
\label{O'cconorYdri action}
\end{eqnarray}
was studied both numerically and perturbative and
the critical point is found to be 
\begin{eqnarray}
\rho_c=\left(\frac{8}{m^2+\sqrt{2}-1}\right)^{1/4}. 
\label{O'cconorYdri critical value}
\end{eqnarray}
 By redefining the field and by identifying parameters as
\begin{eqnarray}
A_i=g^{1/2}N^{1/4}X_i,\qquad 
m^2\sim N^2, \qquad
\alpha\sim g^{1/2}N^{1/4}\rho,
 \label{redef of field}
\end{eqnarray}
 we obtain (\ref{steinacker's action}) from (\ref{O'cconorYdri action}) up to $O(1)$ factors.
With this identification, the scaling of the critical coupling becomes 
\begin{eqnarray}
\frac{1}{g_c^2}\sim (N^{1/4}\rho_c\alpha^{-1})^4\sim N, 
\end{eqnarray}
which agrees with the rough estimation just below (\ref{steinacker_collapse}).
Therefore it finally follows that we cannot formulate 
$D=2$ NCYM with an adjoint scalar using (\ref{steinacker's action}). 

However, the generalized model (\ref{O'cconorYdri action}) has another NCYM limit.
To prevent fuzzy sphere from collapsing in the continuum limit ($g^2$ fixed and $\alpha \sim 1/ \sqrt N$) we have to scale $\frac{1}{g_c^2}\lesssim O(1)$.
To realize this scaling with the redefinition of the field  and identification of $\alpha$ shown in (\ref{redef of field}),
we have to scale $m$ as
\begin{eqnarray}
m^2\gtrsim N^3, \label{new scaling of m}
\end{eqnarray}
instead of $N^2$.
Since last two terms in (\ref{O'cconorYdri action}) are rewritten as  
\begin{eqnarray}
N\cdot \frac{2m^2}{N^2-1}
Tr\left(
X_i^2
-
\frac{N^2-1}{4}\rho^2
\right)^2
+{\rm const}, 
\end{eqnarray}
they suppress the fluctuation perpendicular to fuzzy sphere.   
Therefore, an adjoint scalar, which corresponds to this direction decouples and 
we obtain $D=2$ pure NCYM with the scaling (\ref{new scaling of m}). 

Before closing this subsection let us remark on the subtlety in the above argument. The bound (\ref{O'cconorYdri critical value}) is obtained by calculating the free energy of the rescaled fuzzy sphere. 
However, if the value of $m$ is extremely large, the instability (if exists) 
cannot be captured in this way, because collapse without changing 
the value of $A_i^2$ is more economical.
If adjoint scalars decouple, the situation is analogous to 
the case of TEK.   
In $D=2$ TEK, fuzzy $T^2$ does not collape. 
Therefore in the case of fuzzy $S^2$, we do not expect this kind of instability. 
On the other hand, in the case of
fuzzy $S^2\times S^2$ or fuzzy $\mathbb{C}P^2$ with adjoint scalar potentials 
\cite{GS05,DY06}, we expect this instability similarly to fuzzy $T^4$ in $D=4$ TEK and 
hence NCYM on fuzzy ${\mathbb R}^4$ cannot be obtained\footnote{There 
are models in which adjoint scalars are 
dropped by hand. We expect the situation is the same \cite{GS05,STS07}.}.
It is desirable to check it directly with Monte-Carlo simulation. 
\subsection{Other limits}\label{otherlimit}
\hspace{0.51cm}
So far, we considered  
the continuum limit with the noncommutativity parameter fixed 
and showed that most of bosonic models have instability. 
In this subsection, we discuss other possible limits.
For concreteness we consider $D=4$ TEK model. 

First, let us consider the case in which the fuzzy torus does not collapse. 
The noncommutativity parameter is expressed as  
\begin{eqnarray}
\theta \sim \frac{1}{\sqrt{N}a^2} \sim \frac{\Lambda^2}{\sqrt{N}}.  
\end{eqnarray}
To prevent the fuzzy torus from collapsing, the momentum cutoff 
must be large enough so that 
\begin{eqnarray}
\log\Lambda\sim\frac{1}{g^2}\sim
\Lambda^{-4}\beta N\gtrsim \Lambda^{-4}N^2.
\label{loglambda} 
\end{eqnarray}
On the other hand, to keep the volume of noncommutative space 
$a\sqrt{N}\sim\sqrt{N}/\Lambda$ nonzero, $\Lambda$ cannot be 
so large:  
\begin{eqnarray}
\Lambda\lesssim\sqrt{N}. 
\label{lambda}
\end{eqnarray} 
The only solution to the above constraints (\ref{loglambda}) 
and (\ref{lambda}) is 
\begin{eqnarray}
\theta\sim\sqrt{N}, \qquad\Lambda\sim\sqrt{N},  
\end{eqnarray}
up to $\log\Lambda$ corrections.
In this limit, noncommutativity length $\theta^{-1}$ goes to zero and spacetime volume is fixed. This limit has been studied in many references. This limit is of interest as an alternative to 
lattice gauge theory, because it might 
provide simpler way to introduce chiral fermions \cite{GP95WW97AIN02}.   

Next let us consider the case that fuzzy torus does collapse. 
From D-brane point of view, it just means D-brane collapses to lower dimensional configuration. From NCYM perspective  
this limit seems not to have a sensible continuum limit because there is no extended direction.  
In \cite{BNSV06} slightly different model with two commutative 
and two noncommutative dimensions has been studied numerically. 
In that case two noncommutative dimensions collapse similarly to 
our case, but numerical results suggest that there is a continuum limit 
with two commutative noncompact directions and two compact, finite size 
``noncommutative" directions. Such models would be interesting as a toy model for compactification mechanism in matrix models. 
\section{Supersymmetric matrix model and noncommutative super Yang-Mills}
\label{sec:NCSYM}
\hspace{0.51cm}
In the previous section, we have discussed various matrix model formulations of bosonic NCYM. In this section we explain the formulation of noncommutative {\it super} Yang-Mills (NCSYM).
For this purpose we introduce 
matrix models with an approximate supersymmetry and 
perturbatively discuss stability of noncommutative spaces.  

Let us consider the IKKT-like matrix model \cite{IKKT96} with a 
cubic Myers term 
\begin{eqnarray}
S=\frac{1}{g^2}Tr\left(
-\frac{1}{4}[A_\mu,A_\nu]^2
+
\frac{2i}{3}\alpha f^{\mu\nu\rho}A_\mu A_\nu A_\rho
-
\frac{1}{2}\bar{\psi}\Gamma^\mu[A_\mu,\psi]
\right), 
\end{eqnarray}
where $A_{\mu}$ and $\psi$ are bosonic and fermionic Hermitian $SU(N)$ matrices,
Greek indices run from $1$ to $d$ $(d=4,6,10)$,  
$\psi$ has a spinor index and 
$\Gamma^\mu$ is the $SO(d)$ Gamma matrix. 
$f^{\mu\nu\rho}$ is the structure constant of a Lie group 
whose rank $r$ is less than $d$.
Except for the cubic Myers term, we can obtain this action from 
$D$-dimensional ${\cal N}=1$ $SU(N)$ super Yang-Mills by dimensional reduction.
  
Since numerical simulations for these matrix models are difficult, except for 
$d=4$ case \cite{AANN05} due to the notorious sign problem, it is difficult to 
discuss stability of backgrounds nonperturbatively. Hence, we provide only perturbative arguments, which works perfectly well for bosonic models. 
The perturbative argument is carried out similarly to the case of bosonic analog of IKKT-like matrix model with a cubic Myers term.

Although this model is not supersymmetric\footnote{
$d=r=3$ model with the cubic term is supersymmetric \cite{IKTW}. 
However, in this case the finiteness of partition function is not 
known for generic $N$
\cite{AW01}. },  
the noncommutative background can be stabilized for any value 
of $\alpha$.  
At large $\alpha$ it is stable because 
potential barrier is very high and, furthermore, fluctuations are suppressed due to approximate supersymmetry. 
At small $\alpha$, it can be stabilized
since this model is almost supersymmetric (at $\alpha=0$ the 
supersymmetry recovers) and this background is almost BPS. 

As a concrete example, let us take $d=10$ and $f_{\mu\nu\rho}$ to be 
the structure constant of $SU(2)\times SU(2)$. 
Fuzzy $S^2\times S^2$ (\ref{classical solution_S2*S2}) is one of the classical solutions for it. 
(Indeed, there is a subtlety for this background. 
We will discuss it in \S\ref{sec:S2*S2}.)
Quantum corrections to this background is calculated in \cite{IKTT03}.
Here we consider the deformation in radial direction only as we have explained in \S\ref{sec:cubic term}.
Up to the leading order of $1/N$, the tree level action $\Gamma_{tree}$
and the one-loop correction $\Gamma_{1-loop}$ 
for rescaled fuzzy sphere $P_\mu=(1+\epsilon)\alpha J_\mu$ are 
calculated as
\begin{eqnarray}
\Gamma_{tree}
&=&
\frac{\alpha^4N^2}{4g^2}
\left\{
(1+\epsilon)^4
-
\frac{4}{3}(1+\epsilon)^3
\right\},\nonumber\\
\Gamma_{1-loop}
&=&
N\cdot 2\log 2\cdot
\left(
2
+
\frac{\epsilon^2}{(1+\epsilon)^2}
\right). 
\label{radiative correction to S2*S2}
\end{eqnarray}  
(See Appendix \ref{appendix:oneloop} for derivation.)
If we scale $\alpha\sim N^{-1/4}$, we have 
\begin{eqnarray}
\Gamma_{tree}\sim -Ng^{-2},
\qquad
\Gamma_{1-loop}\sim N.   
\label{approxcorrection}
\end{eqnarray}
The matrix model we are considering here is 
expected to realize $D=4$ ${\cal N}=4$ NCSYM  
in the continuum limit 
and then the coupling $g$ does not run at one-loop level in this limit.  

From (\ref{approxcorrection}), the one-loop correction 
is smaller than tree-level action 
provided that $g^2$ is sufficiently small. 
We also notice that $n$-loop effect is  
$O(N(g^2/(\alpha^4 N))^{n-1})= Ng^{2(n-1)}$ as a result of approximate SUSY and higher loop effects are negligible in this case
\cite{IKTT03}. 
We therefore see that the classical minimum $\epsilon=0$ survives after 
taking into account quantum corrections and 
we can expect that the fuzzy $S^2\times S^2$ does not collapse.  
On the other hand, at strong coupling it might collapse.  
To overcome this difficulty, it is probably useful to consider 
a supersymmetric deformation in \cite{Bonelli02,FuzzyTorusInHermitianModel}. 

In the above construction using the matrix model with the cubic Myers term, 
only extended supersymmetry can be realized.
In order to construct ${\cal N}=1$ 
NCSYM, supersymmetric generalization of TEK would be necessary. 
\subsection{Subtlety for $S^2\times S^2$ case}\label{sec:S2*S2}
\hspace{0.51cm}
In this subsection we discuss a subtlety for fuzzy $S^2\times S^2$ background. 

Because fuzzy $S^2$ has smaller free energy,  
fuzzy $S^2\times S^2$ is not stable; one of the $S^2$ can shrink, while 
the other expands \cite{S2*S2instability}. 
We notice that $SU(2)\times SU(2)$ is preserved in this proces and
that this instability 
cannot be read off from the one-loop effective action 
(\ref{radiative correction to S2*S2}). 
To avoid this instability, 
we should use four-dimensional fuzzy manifolds with higher symmetry, 
e.g. the fuzzy ${\mathbb C}P^2$. 
${\mathbb C}P^2$ can be stable since the symmetry must be broken 
during the transition to $S^2$. 
The effective action 
does not change qualitatively \cite{KKT05CP2} and  
we can realize $D=4$ ${\cal N}=4$ NCSYM using ${\mathbb C}P^2$.  

It is difficult to realize $D=4$ ${\cal N}=2$ NCSYM using matrix
model formulation because of the instability of 
fuzzy $S^2\times S^2$. 
Naively, if we add a cubic Myers term to $d=6$ supersymmetric matrix model as above,
four-dimensional ${\cal N}=2$ NCSYM is expected to be realized in the continuum limit. 
In this case, the coupling runs as $g^{-2}\sim\log N$, and hence 
the background is stable. 
However, to realize four-dimensional compact fuzzy space 
with $6$ matrices in this model, we need to use $S^2\times S^2$. 
It is necessary to fix the radii somehow, for example by quenching the background 
or adding a small potential term to the adjoint scalars, while keeping 
the continuum theory unchanged. 
Instead of fuzzy $S^2\times S^2$ the fuzzy $S^4$ might be useful.  
To make fuzzy $S^4$ a classical solution, we have to add quintic Myers term. However, it is difficult to discuss the stability because perturbative calculation is not valid. 
 \section{Conclusions and discussions}
 \hspace{0.51cm}
In this paper, we studied the stability of noncommutative spaces in 
several matrix models and discussed whether or not 
they provide nonperturbative formulation of noncommutative Yang-Mills theory (NCYM). 
It turns out that most of matrix model formulations of 
{\it bosonic} NCYM on fuzzy ${\mathbb R}^D$ do not work. 
The only exception we found is $D=2$ pure NCYM. 
In the context of D-branes dynamics, those not realized 
correspond to false vacua. 
This might be a negative conclusion if one regards  
NCYM itself as a UV complete theory.  
However, as an effective description for a D-brane system, 
these bosonic NCYM correctly reproduce the instability of the system. 
According to \cite{VanRaamsdonk01},  
large one-loop correction to free energy, which leads instability of NCYM, is 
due to UV/IR mixing. Hence by eliminating UV/IR mixing somehow,
we expect that NCYM be stabilized.

On the other hand, 
noncommutative {\it super} Yang-Mills(NCSYM) on fuzzy $\mathbb{R}^4$
with extended supersymmetry 
can be formulated using a supersymmetric matrix model deformed by a cubic Myers term. 
At least, as we have see above, $D=4$ $\mathcal{N}=4$ NCSYM in weak coupling 
is realized using this formulation.
Also in certain non-supersymmetric model with adjoint fermions, $\mathbb{Z}_N$ symmetry is not broken \cite{Kovtun:2007py}. Then combining it with twist prescription a certain non-supersymmetric NCYM will be obtained.

Here we comment on the formulations of NCSYM at finite temperature. 
For this purpose, 
we consider supersymmetric matrix quantum mechanics\footnote{
Monte Carlo simulation for supersymmetric matrix quantum mechanics without 
cubic term has been performed recently \cite{HNT07,CW07}, and 
incorporation of a cubic term \cite{BMN02} will be straightforward.  
Thermodynamical property of fuzzy sphere in bosonic model is 
studied in \cite{KNT07}. 
} with
Euclidean time direction compactified and antiperiodic boundary condition 
for fermionic variables imposed. At high temperature, fermionic modes decouples and 
the theory becomes essentially bosonic. Therefore, we can expect that 
noncompact fuzzy space cannot be constructed in the high temperature limit.   
Whether NCSYM at nonzero temperature exists or not 
is a subtle problem and numerical simulation 
along the line of \cite{HNT07,CW07} will be necessary. 

Though we have discussed matrix models formulation only in this paper, 
there is another candidate for nonperturbative formulation of bosonic NCYM 
\cite{AMNS00Apr}. 
However, it seems to share the same problem with the matrix model formulation considered in this paper. 
In \cite{AMNS00Apr} NCYM is mapped to a lattice gauge theory with 
twisted boundary condition. In the continuum limit with noncommutativity 
parameter fixed, however, corresponding lattice gauge theory goes to zero volume 
and essentially reduces to the TEK model (see Appendix \ref{sec:lattice}.) 

Of course, the pathology discussed above does not prevent us from nonperturbative formulations of non-gauge theories on noncommutative spaces using matrix models. 
For example, scalar field theories are well defined and 
we can numerically analyze them using matrix model formulations \cite{BHN04}\cite{Panero06}.
We also notice, as explained in \S\ref{otherlimit}, we can take 
the ``commutative" limit of NCYM, in which the noncommutativity length 
$\theta^{-1/2}$ goes to zero. 
Therefore, one may still regard NCYM as an alternative to the lattice construction for gauge theories on commutative spaces.  

In the end, we comment on the recent progress in TEK and its relation to matrix formulation of bosonic NCYM.
Since collapse of fuzzy sphere in TEK model 
is nothing but the breakdown of ${\mathbb Z}_N$ 
symmetry (original motivation for TEK is to keep this symmetry unbroken), 
construction for bosonic NCYM is tightly related to 
a modification of Eguchi-Kawai model \cite{EK82} such that 
${\mathbb Z}_N$ does not break and large-$N$ reduction works. 
Historically two options have been studied. One is TEK, which works 
fine at $D=2$ but turns out to fail at $D=4$. 
Another one is the quenched Eguchi-Kawai model (QEK)\cite{QEK}, in which 
commutative and extended background is ``quenched" by hand. 
Naively by combining twist and 
quench prescriptions, i.e. by fixing {\it noncommutative} background by hand, 
NCYM seems to be realized. However, it does not seem to work. Indeed, recently 
it was argued that QEK does not work due to the following reason \cite{BS08}. 
In QEK, unitary link variables $U_\mu$'s are constrained to be 
$V_\mu e^{iP_\mu} V_\mu^\dagger$, where $P_\mu= {\rm diag}(p_\mu^1,\cdots,p_\mu^N)$ 
is fixed, $V_\mu$'s are unitary matrices and $p_\mu^i$'s are distributed 
uniformly in ${\mathbb R}^4$. Naively one expects $V_\mu$'s fluctuate around 
$\textbf{1}_N$ and, therefore, $\mathbb{Z}_N$ is not broken. However, what 
actually happens is that $V_\mu$'s become certain permutation matrices, so that     
quenched momenta are ``locked" \cite{BS08} and free energy becomes smaller. 
Intuitively, this result implies, even if the background is quenched by hand, 
$V_\mu$ can get a nontrivial VEV and an essentially 
different background emerges. 

The same can take place also when we quench the noncommutative background. Such a subtlety does not 
exist in supersymmetric case, and $D=4$ ${\cal N}=2$ NCSYM would be realized 
by quenching fizzy $S^2\times S^2$ background. 

Recently a new deformation to Eguchi-Kawai model was proposed in \cite{UY08}. 
They added potential terms for Wilson lines to prevent ${\mathbb Z}_N$ from 
breakdown and argued that the additional terms do not contribute 
in the continuum limit. If it really works, by combining this method with 
the twist prescription, bosonic NCYM might be realized. 
Then, it would be interesting to understand the meaning of the deformation in the context of D-brane dynamics. 

In this paper, we assumed the running of the coupling constant is determined by 
one-loop beta function when we discuss the case of $D=4$. 
However, renormalizability of the NCYM is of course controversial. It will be better if we can determine the running more rigorously, for example, by calculating correlation functions using numerical simulations.  
\begin{center}
\textbf{Acknowledgments}
\end{center}
The authors would like to thank Hikaru Kawai, Yusuke Kimura, Lorenzo Mannelli, 
Jun Nishimura,  
Yuya Sasai and Hidehiko Shimada for stimulating discussions and comments.  
T.~A. and T.~H. would like to thank the Japan Society for the Promotion of Science
for financial support. 
\appendix
\section{Derivation of one-loop effective action in supersymmetric matrix model}
\label{appendix:oneloop}
\hspace{0.51cm}
Let us expand the action  
\begin{eqnarray}
S=\frac{1}{g^2}Tr\left(
-\frac{1}{4}[A_\mu,A_\nu]^2
+
\frac{2i}{3}\alpha f^{\mu\nu\rho}A_\mu A_\nu A_\rho
-
\frac{1}{2}\bar{\psi}\Gamma^\mu[A_\mu,\psi]
\right)
\end{eqnarray}
about the rescaled fuzzy sphere 
\begin{eqnarray}
P_\mu=(1+\epsilon)\alpha J_\mu. 
\end{eqnarray}
At tree level, we have 
\begin{eqnarray}
\Gamma_{tree}
=
\frac{\alpha^4}{g^2}Ns(s+1)
\left\{
(1+\epsilon)^4
-
\frac{4}{3}(1+\epsilon)^3
\right\}
\sim
\frac{\alpha^4N^2}{4g^2}
\left\{
(1+\epsilon)^4
-
\frac{4}{3}(1+\epsilon)^3
\right\}, 
\end{eqnarray}
where $N=(2s+1)^2$. 
Then, the one-loop effective action is \cite{IKKT96} 
\begin{eqnarray}
\Gamma_{1loop}
=
\frac{1}{2}Tr\log\left(
\delta_{\mu\nu}
-
\frac{\epsilon f^{\mu\nu\rho}}{1+\epsilon}\frac{ad J_\rho}{(ad J)^2}
\right)
-
\frac{1}{4}Tr\log\left\{
\left(
1
+
\frac{i}{2}
\Gamma^{\mu\nu}f^{\mu\nu\rho}
\frac{ad J_\rho}{(ad J)^2}
\right)
\frac{1+\Gamma^{11}}{2}
\right\}. 
\nonumber\\
\end{eqnarray}
To leading order in $N$, we have 
\begin{eqnarray}
\Gamma_{1loop}
&=&
\frac{1}{2}Tr\left\{
\frac{1}{2}
\left(
\frac{\epsilon f^{\mu\nu\rho}}{1+\epsilon}\frac{ad J_\rho}{(ad J)^2}
\right)^2
\right\}
-
\frac{1}{4}Tr\left\{
-
\frac{1}{2}
\left(
\frac{i}{2}
\Gamma^{\mu\nu}f^{\mu\nu\rho}
\frac{ad J_\rho}{(ad J)^2}
\right)^2
\frac{1+\Gamma^{11}}{2}
\right\}
\nonumber\\
&=& 
\left(
2
+
\frac{\epsilon^2}{(1+\epsilon)^2}
\right)Tr 
\frac{1}{(ad J)^2}
\nonumber\\
&=&
N\cdot 2\log 2\cdot
\left(
2
+
\frac{\epsilon^2}{(1+\epsilon)^2}
\right), 
\end{eqnarray}
where we have used \cite{IKTT03}
\begin{eqnarray}
Tr 
\frac{1}{(ad J)^2}
=
Tr 
\frac{1}{(ad J^{(s)}\otimes 1)^2 +(1\otimes ad J^{(s)})^2 }
=
\sum_{j=1}^{2s}
\sum_{j^\prime=1}^{2s}
\frac{(2j+1)(2j^\prime+1)}{j(j+1)+j^\prime(j^\prime+1)}
\simeq
2N\log 2. 
\nonumber\\
\end{eqnarray}
\section{Lattice formulation}\label{sec:lattice}
\hspace{0.51cm}
Lattice regularization \cite{AMNS00Apr} relates commutative $U(N)$ lattice gauge theory on twisted torus to a ``lattice regularization" of $U(1)$ NCYM on periodic fuzzy torus. 
Basically this relation is as a result of the fact that the Morita equivalence holds at lattice level.

For simplicity, we consider the $D=4$ $U(N)$ gauge theory on a rectangular four-torus 
with period $L$.
The action is 
\begin{eqnarray}
S=-\frac{1}{g^2} \sum_x \sum_{\mu \neq \nu}tr \left[ U_\mu(x) U_\nu (x+a \hat\mu) U_\mu(x+a \hat\nu)^\dagger U_\nu(x)^\dagger \right], \label{action of TYM}
\end{eqnarray}
where $U_\mu$ are unitary matrices which correspond to $U(N)$ gauge fields. 
They satisfy twisted boundary condition 
\begin{eqnarray}
U_\mu(x+l\hat{\nu})=\Gamma_\nu U_\mu \Gamma_\nu^\dagger, 
\end{eqnarray}
where $\Gamma_\nu$ are twist-eaters appeared in \S\ref{sec:TEK}. 

We now introduce a map $\hat \Delta(x)$ between lattice fields $U_\mu (x)$ and operators $\hat U_\mu$ as
\begin{eqnarray}
\hat U_\mu = \sum_x \hat\Delta(x) U_\mu(x)
\end{eqnarray}
where the mapping function $\hat\Delta(x)$
is defined as
\begin{eqnarray}
\hat \Delta(x) = \left(\frac{l}{a} \right)^N \sum_{m^i \in 
\mathbb{Z}/n}
\left(\prod^4_{i=1}e^{i k_a (\hat x_a - x_a)}  \right),
\end{eqnarray}
where $k_a$ is a momentum $k_a = 2 \pi m_a /l$ and $n$ is a integer $n=l/a$.

In order to relate operators $\hat{U}_{\mu}$ to noncommutative $U(1)$ gauge fields,  
we now introduce another mapping function $\hat \Delta'(x')$ defined as
\begin{eqnarray}
\hat \Delta '(x')
=\left(\frac{l'}{\epsilon} \right)^N
e^{-\pi i\sum_{a<b}m_a\Theta_{ab} m_b}
\sum_{m^a \in \mathbb{Z}/n'}
\left(\prod^4_{a=1}e^{i k'_a (\hat x_a - x'_a)}  \right),
\end{eqnarray}
where $l'$ = $l\sqrt{N}$, $k'_{a}=2 \pi m_a /l'$, $n'=l'/a$ and 
\begin{eqnarray}
\Theta_{ab}
=
\left(
\begin{array}{cccc}
0 & \Theta & 0 & 0\\
-\Theta & 0 & 0 & 0\\
0 & 0 & 0 & \Theta\\
0 & 0 & -\Theta & 0  
\end{array}
\right)_{ab},
\qquad
\Theta=\frac{1}{\sqrt{N}}. 
\end{eqnarray}
We have used primed quantities to represent those on a lattice corresponding to $\Delta'$.
This $\hat\Delta'(x')$ maps the noncommutative lattice fields to operators whose dimensionless noncommutativity parameters is $\Theta$.
Because of the twist boundary condition of $U_\mu(x)$ the operator $\hat U_\mu$ have another expansion using $\hat \Delta'(x')$,
\begin{eqnarray}
\hat U_\mu = \sum_{x'} \hat\Delta'(x') U'_\mu(x'),
\end{eqnarray}
where $U'_\mu(x')$ are noncommutative $U(1)$ gauge fields which live in periodic torus whose size is $l'$ and the dimensionless noncommutativity parameter is $\Theta$.

Now we gain a map from $U(N)$ gauge fields $U_\mu(x)$ on a twisted commutative torus to  the noncommutative $U(1)$ gauge fields $U'_\mu(x')$ on a periodic fuzzy torus.
Indeed the action (\ref{action of TYM}) is rewritten in terms of $U'_\mu(x')$ as
\begin{eqnarray}
S=-\frac{1}{{g'}^2} \sum_{x'} \sum_{\mu \neq \nu}tr \left[ U'_\mu(x')\star U'_\nu (x'+a \hat\mu)\star U'_\mu(x'+a \hat\nu)^\dagger \star U'_\nu(x') ^\dagger \right],
\end{eqnarray}
where 
\begin{eqnarray}
{g'}^2 = N g^2.
\end{eqnarray}

Dimensionful noncommutativity parameter, which appears in commutators of coordinates is given by
\begin{eqnarray}
\theta
=
\Theta\cdot\frac{l^{\prime 2}}{2\pi}
=
\frac{l^2\sqrt{N}}{2\pi}. 
\end{eqnarray}

Now let us consider the limit which leads to fuzzy ${\mathbb R}^4$  with finite value of $\theta$. 
To fix $\theta$, we have to take 
\begin{eqnarray}
l\sim N^{-1/4},  
\end{eqnarray}
that is, we have to take infinitely small twisted torus and 
the model essentially reduces to TEK.  
Therefore it is plausible that the center symmetry $U(1)$ breaks down.
This means that the fuzzy torus collapses and we cannot realize 
fuzzy $\mathbb{R}^4$ which is expected to appear as a tangent space
of the torus.




\begin{thebibliography}{3}


\bibitem{SW99}
  N.~Seiberg and E.~Witten,
  JHEP {\bf 9909} (1999) 032; hep-th/9908142.
 
\bibitem{BFSS96}
 T.~Banks, W.~Fischler, S.~H.~Shenker and L.~Susskind,
 Phys.\ Rev.\  D {\bf 55} (1997) 5112;
 hep-th/9610043.

\bibitem{IKKT96}
 N.~Ishibashi, H.~Kawai, Y.~Kitazawa and A.~Tsuchiya,
 Nucl.\ Phys.\  B {\bf 498} (1997) 467;
 hep-th/9612115.
\bibitem{CDS97}
 A.~Connes, M.~R.~Douglas and A.~S.~Schwarz,
 JHEP {\bf 9802} (1998) 003;
 hep-th/9711162.  
 
\bibitem{AIIKKT99}
 H.~Aoki, N.~Ishibashi, S.~Iso, H.~Kawai, Y.~Kitazawa and T.~Tada,
 Nucl.\ Phys.\  B {\bf 565} (2000) 176;
 hep-th/9908141.

\bibitem{Li96}
 M.~Li,
 Nucl.\ Phys.\  B {\bf 499} (1997) 149;
 hep-th/9612222.

 \bibitem{GMS00}
  R,~Gopakumar, S.~Minwalla and A.~Strominger,
  JHEP {\bf 0005} (2000) 020; 
  hep-th/0003160.

\bibitem{GravityInNCYM}
  Y.~Kitazawa and S.~Nagaoka,
  JHEP {\bf 0602} (2006) 001; 
  hep-th/0512204.

	H.~Steinacker,
  JHEP {\bf 0712} (2007) 049; 
  arXiv:0708.2426 [hep-th].

\bibitem{SteinackerMatter}
   H.~Steinacker
   arXiv:0806.2032 
  
   
\bibitem{EK82}
  T.~Eguchi and H.~Kawai,
  Phys.\ Rev.\ Lett.\  {\bf 48} (1982) 1063.


\bibitem{QEK} 
  G.~Parisi,
  Phys.\ Lett.\  B {\bf 112} (1982) 463.
  
   G.~Bhanot, U.~M.~Heller and H.~Neuberger,
  Phys.\ Lett.\  B {\bf 113} (1982) 47. 
  
   D.~J.~Gross and Y.~Kitazawa,
  Nucl.\ Phys.\  B {\bf 206} (1982) 440.
 
\bibitem{GAO82}

 A.~Gonzalez-Arroyo and M.~Okawa,
 Phys.\ Rev.\ D {\bf 27} (1983) 2397.
  \bibitem{EN82}
T.~Eguchi and R.~Nakayama
Phys.\ Lett.\ B {\bf 122} (1983) 59.


  
\bibitem{EmergentGeometryInIKKT}

  M.~Fukuma, H.~Kawai, Y.~Kitazawa and A.~Tsuchiya,
  Nucl.\ Phys.\  B {\bf 510} (1998) 158; 
  hep-th/9705128.	

  H.~Aoki, S.~Iso, H.~Kawai, Y.~Kitazawa and T.~Tada,
  Prog.\ Theor.\ Phys.\  {\bf 99} (1998) 713; 
  hep-th/9802085. 
  
\bibitem{DifferentialOperator}
   M.~Hanada, H.~Kawai and Y.~Kimura,
  Prog.\ Theor.\ Phys.\  {\bf 114} (2006) 1295; 
  hep-th/0508211.
 
   M.~Hanada, H.~Kawai and Y.~Kimura,
  Prog.\ Theor.\ Phys.\  {\bf 115} (2006) 1003; 
  hep-th/0602210.
 
   M.~Hanada,
  Prog.\ Theor.\ Phys.\  {\bf 115} (2006) 1189; 
  hep-th/0606163.
 
   K.~Furuta, M.~Hanada, H.~Kawai and Y.~Kimura,
  Nucl.\ Phys.\  B {\bf 767} (2007) 82; 
  hep-th/0611093.


\bibitem{Maldacena97}
  J.~M.~Maldacena,
  Adv.\ Theor.\ Math.\ Phys.\  {\bf 2} (1998) 231 
  [Int.\ J.\ Theor.\ Phys.\  {\bf 38} (1999) 1113]; 
  hep-th/9711200.
  
\bibitem{Berenstein05}
  D.~Berenstein,
  JHEP {\bf 0601} (2006) 125; 
  hep-th/0507203.

\bibitem{Myers99}
  R.~C.~Myers,
  JHEP {\bf 9912} (1999) 022; 
  hep-th/9910053.

\bibitem{BHN02}
  W.~Bietenholz, F.~Hofheinz and J.~Nishimura,
  JHEP {\bf 0209} (2002) 009; hep-th/0203151.  

\bibitem{MRS99}
 S.~Minwalla, M.~Van Raamsdonk and N.~Seiberg,
  JHEP {\bf 0002},(2000) 020 ;
  hep-th/9912072.
  
  I.~L.~Buchbinder and V.~A.~Krykhtin,
  Int.\ J.\ Mod.\ Phys.\  A {\bf 18} (2003) 3057; 
  hep-th/0207086. 
  
  \bibitem{VanRaamsdonk01}
  M.~Van Raamsdonk,
  JHEP {\bf 0111} (2001) 006; 
  hep-th/0110093. 
  
  A.~Armoni and E.~Lopez,
  Nucl.\ Phys.\  B {\bf 632} (2002) 240; 
  hep-th/0110113.  
  
  
\bibitem{AHHI07}  
  T.~Azeyanagi, M.~Hanada, T.~Hirata and T.~Ishikawa,
  JHEP {\bf 0801} (2008) 025; 
  arXiv:0711.1925 [hep-lat].  
   
\bibitem{TV06}
 M.~Teper and H.~Vairinhos,
   Phys.\ Lett.\  B {\bf 652} (2007) 359; 
  arXiv:hep-th/0612097.
  
\bibitem{AMNS99}
 J.~Ambjorn, Y.~M.~Makeenko, J.~Nishimura and R.~J.~Szabo,
 JHEP {\bf 9911} (1999) 029;
 hep-th/9911041.

\bibitem{HNT98}
 T.~Hotta, J.~Nishimura and A.~Tsuchiya, 
  Nucl.\ Phys.\  B {\bf 545} (1999) 543; 
  hep-th/9811220. 
 
\bibitem{ABNN04S4}
 T.~Azuma, S.~Bal, K.~Nagao and J.~Nisimura, 
 JHEP {\bf 0407} (2004) 0666; 
 hep-th/0405096. 
 
\bibitem{InstabilityOfFuzzySphere} 
  T.~Azuma, S.~Bal, K.~Nagao and J.~Nishimura,
  JHEP {\bf 0405} (2004) 005; 
  hep-th/0401038.
  
  T.~Azuma, S.~Bal, K.~Nagao and J.~Nishimura,
  JHEP {\bf 0605} (2006) 061; 
  hep-th/0405277. 
  
   T.~Azuma, K.~Nagao and J.~Nishimura,
  JHEP {\bf 0506} (2005) 081; 
  hep-th/0410263.
  
    T.~Azuma, S.~Bal, K.~Nagao and J.~Nishimura,
  JHEP {\bf 0509} (2005) 047; 
  hep-th/0506205.
 
\bibitem{Steinacker03}
  H.~Steinacker,
  Nucl.\ Phys.\  B {\bf 679} (2004) 66; 
  hep-th/0307075.
 
\bibitem{OY06}
  D.~O'Connor and B.~Ydri,
  JHEP {\bf 0611} (2006) 016; 
  hep-lat/0606013. 
 
\bibitem{DY06}

  R.~Delgadillo-Blando and B.~Ydri,
  JHEP {\bf 0703} (2007) 056; 
  hep-th/0611177. 
 
  D.~Dou and B.~Ydri,
  Nucl.\ Phys.\  B {\bf 771} (2007) 167; 
  hep-th/0701160.
 
\bibitem{GS05}
  H.~Grosse and H.~Steinacker,
  Nucl.\ Phys.\  B {\bf 707} (2005) 145; 
  hep-th/0407089.
  
\bibitem{STS07}
  H.~Steinacker and R.~J.~Szabo,
  Commun.Math.Phys. {\bf 278} (2008) 193;
  hep-th/0701041.

\bibitem{GP95WW97AIN02}
 H.~Grosse and P.~Presnajder, 
 Lett.Math.Phys. {\bf 33} (1995) 171;


 U.~Carow-Watamura and S.~Watamura,
Commun.Math.Phys. {\bf 183} (1997) 365;
hep-th/9605003.

H.~Aoki, S.~Iso and K.~Nagao,
Phys.Rev. D{\bf 67} (2003) 085005;
hep-th/0209223. 

   \bibitem{BNSV06}
 W.~Bietenholz, J.~Nishimura, Y.~Susaki and J.~Volkholz,
 JHEP {\bf 0610} (2006) 042;
 hep-th/0608072.
\bibitem{AANN05}
  K.~N.~Anagnostopoulos, T.~Azuma, K.~Nagao and J.~Nishimura,
  JHEP {\bf 0509} (2005) 046; hep-th/0506062.  

\bibitem{IKTW}    
  S.~Iso, Y.~Kimura, K.~Tanaka and K.~Wakatsuki,
  Nucl.\ Phys.\  B {\bf 604} (2001) 121; 
  hep-th/0101102.  

\bibitem{AW01}
  P.~Austing and J.~F.~Wheater,
  JHEP {\bf 0104} (2001) 019; hep-th/0103159. 
 
  P.~Austing and J.~F.~Wheater,
  JHEP {\bf 0311} (2003) 009; hep-th/0310170.
  
  D.~Tomino,
  JHEP {\bf 0401} (2004) 062; hep-th/0309264
  
\bibitem{IKTT03}
  T.~Imai, Y.~Kitazawa, Y.~Takayama and D.~Tomino,
  Nucl.\ Phys.\  B {\bf 679} (2004) 143; 
  hep-th/0307007.
  
  
  
 \bibitem{Bonelli02}
  G.~Bonelli,
  JHEP {\bf 0208} (2002) 022; 
  hep-th/0205213.
  
  P.~Austing,
  arXiv:hep-th/0108128.
  
\bibitem{FuzzyTorusInHermitianModel}
  M.~Unsal,
  JHEP {\bf 0512} (2005) 033; 
  hep-th/0409106. 
  
    H.~Shimada,
  arXiv:0804.3236 [hep-th].

  
\bibitem{S2*S2instability}
  S.~Bal, M.~Hanada, H.~Kawai and F.~Kubo,
  Nucl.\ Phys.\  B {\bf 727} (2005) 196
  [arXiv:hep-th/0412303].

  H.~Kaneko, Y.~Kitazawa and D.~Tomino,
  Nucl.\ Phys.\  B {\bf 725} (2005) 93; 
  arXiv:hep-th/0506033.

\bibitem{KKT05CP2}
 H.~Kaneko, Y.~Kitazawa and D.~Tomino, 
 Phys.\ Rev.\ D{\bf 73} (2006) 660011;    
hep-th/0510263. 


\bibitem{Kovtun:2007py}
  P.~Kovtun, M.~Unsal and L.~G.~Yaffe,
  JHEP {\bf 0706}, 019 (2007); 
  hep-th/0702021.


\bibitem{HNT07}
  M.~Hanada, J.~Nishimura and S.~Takeuchi,
   Phys.\ Rev.\ Lett.\  {\bf 99} (2007) 161602; 
  arXiv:0706.1647 [hep-lat].

  K.~N.~Anagnostopoulos, M.~Hanada, J.~Nishimura and S.~Takeuchi,
   Phys.\ Rev.\ Lett.\  {\bf 100} (2008) 021601; 
  arXiv:0707.4454 [hep-th].

\bibitem{CW07}
  S.~Catterall and T.~Wiseman,
  JHEP {\bf 0712} (2007) 104; 
  arXiv:0706.3518 [hep-lat]. 
  
  S.~Catterall and T.~Wiseman,
  arXiv:0803.4273 [hep-th].

\bibitem{BMN02}
  D.~E.~Berenstein, J.~M.~Maldacena and H.~S.~Nastase,
  JHEP {\bf 0204} (2002) 013; 
  hep-th/0202021.

\bibitem{KNT07}
  N.~Kawahara, J.~Nishimura and S.~Takeuchi,
  JHEP {\bf 0705} (2007) 091; 
  arXiv:0704.3183 [hep-th].
  
\bibitem{AMNS00Apr}
 J.~Ambjorn, Y.~M.~Makeenko, J.~Nishimura and R.~J.~Szabo,
 JHEP {\bf 0005} (2000) 023;
 hep-th/0004147.
  
  
 
\bibitem{BHN04}
  J.~Ambjorn and S.~Catterall,
  Phys.\ Lett.\  B {\bf 549} (2002) 253; 
  hep-lat/0209106.
  
	W.~Bietenholz, F.~Hofheinz and J.~Nishimura,
  JHEP {\bf 0406} (2004) 042; hep-th/0404020. 
  
\bibitem{Panero06}
  M.~Panero,
  SIGMA {\bf 2} (2006) 081; hep-th/0609205.  
  
    J.~Medina, W.~Bietenholz and D.~O'Connor,
  JHEP {\bf 0804} (2008) 041
  [arXiv:0712.3366 [hep-th]].

\bibitem{BS08}
  B.~Bringoltz and S.~R.~Sharpe,
  arXiv:0805.2146 [hep-lat].  
 
\bibitem{UY08}
  M.~Unsal and L.~G.~Yaffe,
  arXiv:0803.0344 [hep-th].


\end{thebibliography}
\end{document}